\pdfoutput=1 

\documentclass[runningheads]{llncs}
\usepackage[table,xcdraw,dvipsnames]{xcolor}

\usepackage{graphicx}
\usepackage{tablefootnote}
\usepackage{amsmath} 
\usepackage{todonotes}

\usepackage{tabularx}  
\usepackage[pdfstartview=XYZ,
bookmarks=true,
colorlinks=true,
linkcolor=blue,
urlcolor=blue,
citecolor=blue,
pdftex,
bookmarks=true,
linktocpage=true, 
hyperindex=true
]{hyperref}

\usepackage{multirow}
\usepackage{orcidlink}
\usepackage{textcomp}
\usepackage{verbatim}
\usepackage{comment}
\usepackage{threeparttable}
\usepackage{makecell}
\usepackage{xspace}
\usepackage{diagbox}
\usepackage{wrapfig}

\usepackage{graphicx}

\usepackage{tabularx}

\usepackage{floatflt}
\usepackage{lipsum}

\begin{document}
\title{OMPGPT: A Generative Pre-trained Transformer Model for OpenMP}
%
%
\author{Le Chen\dag\inst{1} \and
Arijit Bhattacharjee\dag\inst{1} \and
Nesreen Ahmed \inst{2} \and
Niranjan Hasabnis\inst{2} \and
Gal Oren \inst{3} \and
Vy Vo\inst{2} \and
Ali Jannesari\inst{1}
\thanks{\dag: Equal contribution. 1: ISU. 2: Intel. 3: Technion\\
contact: lechen@iastate.edu}}
\authorrunning{Chen et al.}
%
\institute{ }

%
\maketitle              

\setlength{\textfloatsep}{8pt}
\begin{abstract}
Large language models (LLMs)such as ChatGPT have significantly advanced the field of Natural Language Processing (NLP). This trend led to the development of code-based large language models such as StarCoder, WizardCoder, and CodeLlama, which are trained extensively on vast repositories of code and programming languages. 
While the generic abilities of these code LLMs are helpful for many programmers in tasks like code generation, the area of high-performance computing (HPC) has a narrower set of requirements that make a smaller and more domain-specific model a smarter choice. This paper presents OMPGPT, a novel domain-specific model meticulously designed to harness the inherent strengths of language models for OpenMP pragma generation. Furthermore, we leverage prompt engineering techniques from the NLP domain to create Chain-of-OMP, an innovative strategy designed to enhance OMPGPT's effectiveness. Our extensive evaluations demonstrate that OMPGPT outperforms existing large language models specialized in OpenMP tasks and maintains a notably smaller size, aligning it more closely with the typical hardware constraints of HPC environments. We consider our contribution as a pivotal bridge, connecting the advantage of language models with the specific demands of HPC tasks.

\end{abstract}
%


\section{Introduction}
\label{sec:intro}
Recent advancements in transformer-based~\cite{vaswani2017attention} large language models (LLMs) have revolutionized artificial intelligence and machine learning.
These models have shown remarkable performance in natural language processing (NLP) tasks, leading to the development of code-based LLMs such as StarCoder~\cite{li2023starcoder}, WizardCoder~\cite{luo2023wizardcoder}, and CodeLlama~\cite{rozière2023code}, which are specifically designed for programming language tasks. However, applying these models to High-Performance Computing (HPC) tasks presents unique challenges.




\begin{itemize}
    \item Training data diversity: Advanced LLMs like GPT3 and CodeLlama are trained on both natural language (NL) and programming languages (PL), enabling them to interpret NL prompts and generate appropriate PL code. In contrast, models like Starcoder, trained solely on code, struggle with NL prompts and are limited to code generation tasks.
    \item Performance consistency. Using NL as input can lead to variability in LLM outputs due to different phrasing of the same question by users, posing challenges for consistent performance and post-processing.
    \item Output processing. LLMs typically return answers in NL, necessitating additional effort to extract relevant information for practical use.
\end{itemize}


\begin{figure*}[!t]
\centering
\includegraphics[width=\textwidth]{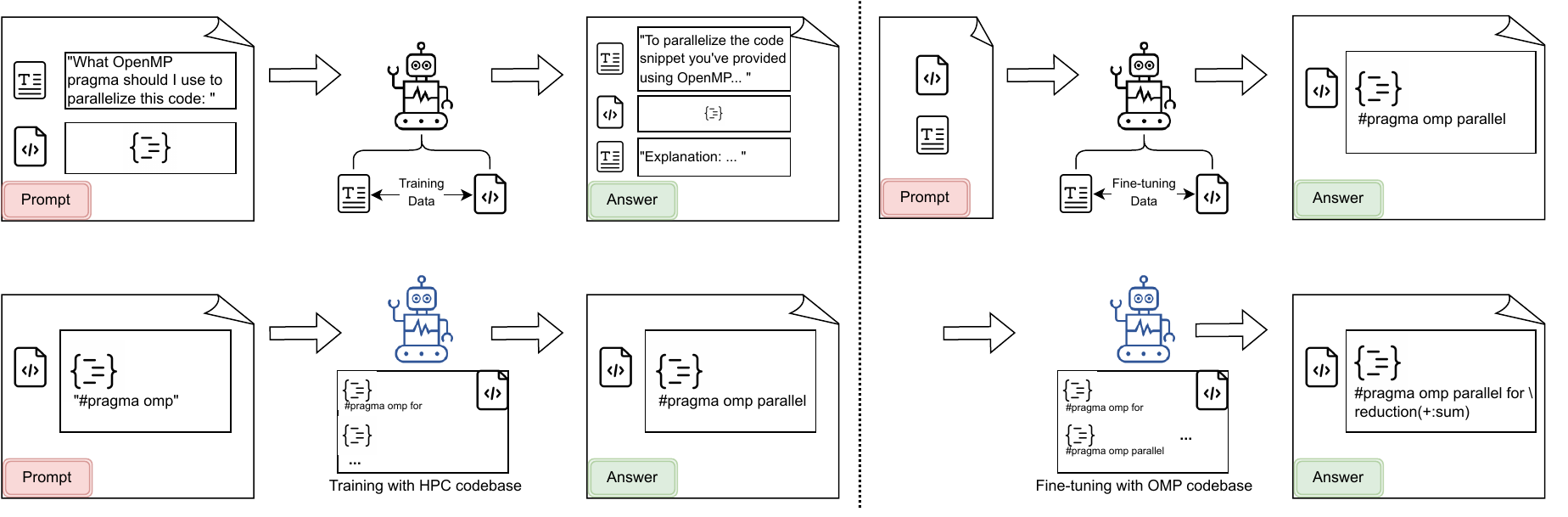} 
\caption{\includegraphics[height=1.2em]{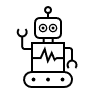} (black): Traditional LLMs require extensive NL and PL training data for generating OpenMP pragmas, leading to an increased complexity and larger model size. Users usually carefully craft prompts and interpret outputs to obtain accurate OpenMP pragmas.
\includegraphics[height=1.2em]{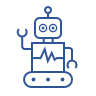} (blue): \textbf{OMPGPT} is tailored for HPC code, with nearly half of its training data being OpenMP code. It aligns OpenMP pragmas with their scope during training to match GPT model instincts. OMPGPT not only addresses the limitations of current LLMs but also benefits from the \textbf{Chain-of-OMP} prompt engineering technique.}
\label{fig:teaser}
\end{figure*}



To address these challenges, we propose the OMPGPT model and the chain-of-OMP prompt engineering method for automatic parallelization via OpenMP pragma generation. As illustrated in Figure~\ref{fig:teaser}, OMPGPT is a domain-specific model trained on an extensive HPC dataset of C and C++ code, converting OpenMP pragma generation into a code generation problem. This approach follows the instinct of LLMs and eliminates the need for training NL data to understand the task objective. The Chain-of-OMP method enhances OMPGPT's performance by incrementally refining prompts with preconditions for OpenMP pragma generation, aligning with the structure of OpenMP pragmas.
The contributions of this work are as follows:

\begin{itemize}
    \item We introduce OMPGPT, a compact 0.76B domain-specific language model (smallest among our baseline models) tailored for OpenMP pragma generation with competitive performance to larger LLMs.
    \item We propose a novel OpenMP clause-based prompting technique, Chain-of-OMP, which enhances OMPGPT by providing targeted hints. 
    \item Our comprehensive evaluation demonstrates OMPGPT's superior performance in OpenMP pragma generation compared to the state-of-the-art models MonoCoder~\cite{kadosh2023domain} and GPT-3.5 and highlights the effectiveness of the Chain-of-OMP method in boosting OMPGPT's capabilities.
\end{itemize}

\section{Background} 
\label{sec-background}

\subsection{Generative Pre-trained Transformers and Code LLMs}
The emergence of Generative Pre-trained Transformers (GPT) has revolutionized Natural Language Processing (NLP) and extended its influence to programming languages, with models like GPT-3.5 capable of generating source code. GPT models are autoregressive, generating text sequentially from left to right, which enables them to produce contextually appropriate responses in natural language tasks. 
This led to the development of Large Language Models (LLMs) for code (Code LLMs), which are able to understand and generate programming code. Code LLMs are trained on extensive datasets containing code snippets and are designed to assist in code completion, quality improvement, and streamlining the software development process. However, challenges remain in ensuring the models can generate contextually and functionally meaningful code.

StarCoder \cite{li2023starcoder} is a 15B parameter model trained for code generation or completion. The training dataset, the Stack \cite{Kocetkov2022TheStack}, has 1 trillion tokens sourced from a large collection of permissively licensed GitHub repositories.
CodeLlama \cite{rozière2023code} is a code-generating LLM based on Llama2 \cite{touvron2023llama} by Meta, specialized for code by training with code-specific datasets. It has parameter sizes of 7B, 13B, and 34B, trained with 500B tokens of code data.
Most code LLMs (e.g., StarCoder) have been trained on raw code data without instruction fine-tuning. However, the recent WizardCoder~\cite{luo2023wizardcoder}, enables Code LLMs with instruction fine-tuning by adapting the Evol-Instruct methods for coding tasks, and it was shown to improve the performance of code generation.

\subsection{LLMs on code-related tasks for HPC}
LLM models show potential in HPC tasks by leveraging their ability to capture complex code patterns and boost efficiency. Although not specifically designed for HPC, their adaptability makes them valuable for code-related HPC challenges~\cite{valerolara2023comparing}. HPC predominantly uses C, C++, and Fortran for low-level control and parallelism optimization capabilities. Focusing on these primary languages allows smaller, more efficient models tailored to specific HPC tasks with direct inputs. Recently, research in applying LLMs to HPC has been emerging. For example, LM4HPC~\cite{chen2023lm4hpc} stands as the first attempt to adapt LLMs to the HPC domain. Building upon this, a subsequent study by Ding et al.~\cite{dingHPCGPTIntegratingLarge2023} introduced a Llama-based Question and Answer model specifically tailored for HPC tasks. Despite these advancements, these early efforts predominantly rely on existing LLMs, indicating a nascent stage in developing HPC-focused language models.

One important HPC task is sequential code parallelization. OpenMP (Open Multi-Processing) \cite{openmp} is the mainstream API that supports multi-platform shared-memory multiprocessing programming in C, C++, and Fortran. It enables developers to write parallel code using multiple cores on a single machine. Various works~\cite{chen2023learning,chen2022multi} have applied machine learning techniques to predict OpenMP pragmas for sequential code parallelization.

\subsection{Prompt Engineering}
Prompt engineering is a key technique in LLMs for crafting input prompts that elicit specific outputs. Chain-of-thought (CoT)~\cite{wei2023chainofthought} is a prominent method where a series of intermediate reasoning steps enhances an LLM's complex reasoning ability. Instead of a single prompt, CoT guides the model through phases of small hints, helping it grasp the deeper meaning of the query. This approach has significantly improved LLM performance in areas like arithmetic and common sense reasoning.



\section{Approach} 
\label{sec-approach}
\subsection{OMPGPT Design}
As highlighted in Section~\ref{sec:intro}, existing Large Language Models (LLMs) encounter significant limitations when applied to HPC tasks, particularly in the realm of OpenMP pragma generation. In designing OMPGPT, we consider several criteria to ensure its suitability and efficacy for OpenMP code generation:

\begin{enumerate}
    \item \textbf{Training Data Relevance}: The quality of the training dataset is essential to any language model. OMPGPT is trained on HPC code in the most common programming languages in the HPC field.
    \item \textbf{Model Compatibility}: The OMPGPT model size is aligned with the hardware configurations prevalent in most HPC clusters.
    \item \textbf{Flexibility and Adaptability}: OMPGPT is designed to be flexible and adaptable, capable of handling a variety of OpenMP pragma generation tasks without the need to craft prompts
    \item \textbf{Performance Efficiency}: OMPGPT is expected to outperform previous small language models and be competitive compared to advanced LLMs.
     \item \textbf{User Accessibility}: Recognizing the varied nature of HPC ecosystems, OMPGPT is designed for a diverse array of users, regardless of their OpenMP knowledge background. 
\end{enumerate}


\subsection{OMPGPT Training \& Inference}
\label{sec:training}
\begin{figure}[htbp]
    \newlength{\figureheight}
    \setlength{\figureheight}{4cm} 

    \begin{minipage}[t]{0.45\textwidth}
        \centering
        \includegraphics[height=\figureheight]{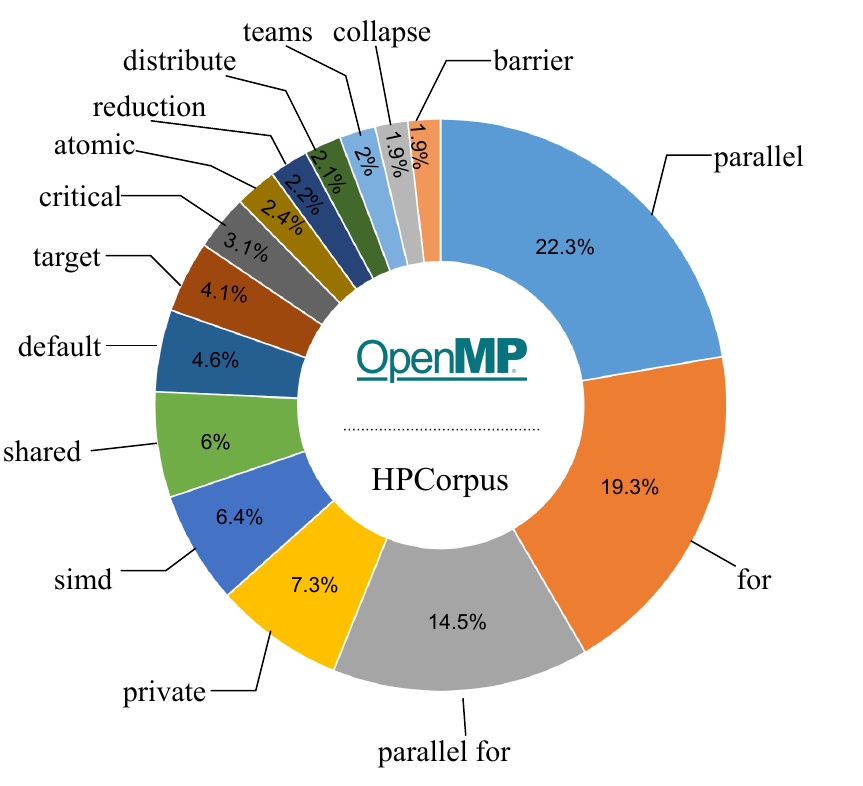}
        \caption{Distribution of OpenMP Pragmas in the HPCorpus OpenMP subset with the top 15 most frequently occurring pragmas.}
        \label{fig:top_15}
    \end{minipage}
    \hfill 
    \begin{minipage}[t]{0.5\textwidth}
        \centering
        \includegraphics[height=\figureheight]{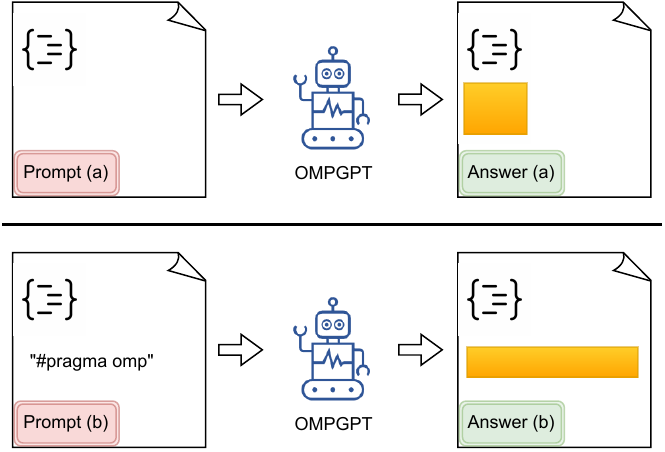}
        \caption{Basic prompt for OMPGPT. (a): prompt for code generation. (b): prompt for OpenMP pragma generation.}
        \label{fig:prompt_base}
    \end{minipage}
\end{figure}

\textbf{Dataset.}  HPCorpus, introduced by Kadosh et al. \cite{kadosh2023quantifying}, is an extensive HPC database derived from GitHub repositories containing C, C++, and Fortran code. Notably, approximately 45\% of the repositories use some form of parallel programming. The primary parallel programming mode is OpenMP, making HPCorpus suitable for training models for OpenMP tasks. 
We use C and C++ code from HPCorpus for our model's training and fine-tuning, with 144,522 C code repositories and 150,481 C++ code repositories, totaling 8,781,759 C code functions and 68,233,984 C++ code functions (a total of 72.39 GB). Figure~\ref{fig:top_15} shows the distribution of OpenMP pragmas in the HPCorpus OpenMP subset used for OMPGPT fine-tuning. We allocate 10\% of the data as a test set for both training and fine-tuning to maintain evaluation integrity.


\noindent
\textbf{Data Processing.}  Our data pre-processing for OMPGPT includes the following:
\begin{enumerate}
    \item \textbf{Trimming natural language}: OMPGPT is 
    trained solely on code. Therefore natural language text such as comments are removed.
    \item \textbf{OpenMP Pragma Positioning}: Our approach, unlike previous studies, places OpenMP pragmas after their scopes, leveraging GPT models' instinct to overcome the constraints of current LLMs. This strategy aligns with the findings of the previous work in~\cite{nichols2023modeling} but diverges by training OMPGPT from scratch to generate OpenMP pragmas, a capability not present in conventional pre-trained models. This unique training methodology tailors OMPGPT to the demands of OpenMP pragma generation and enhances its ability to understand and process the HPC-specific code structures.
    \item \textbf{Filtering code by size}: Following practices used in previous research, such as PolyCoder~\cite{polycoder}, we filter out large code segments (more than 100 tokens or greater than 1 MB) from HPCorpus.
\end{enumerate}


\vspace{-10pt}

\subsubsection{Hardware.}
In all experiments, we utilized a single node on an HPC cluster equipped with a dual-socket AMD EPYC Rome 7402 (24 cores/socket) and 512 GB DDR4-3200 RAM, along with 4 x NVIDIA A100 40GB HBM2e Tensor Core GPUs connected via NVLink3 to each other.


\noindent
\textbf{OMPGPT Training}: We leveraged the GPT-Neo 2.7B \cite{gpt-neo} model architecture as a foundation. However, we have tailored the model to better suit our needs by downsizing the number of layers to 8 while maintaining 32 attention heads per layer. The model features a hidden dimension of 2560. We employ the StarCoder tokenizer for tokenization, which utilizes a vocabulary comprising 49,152 tokens. This configuration results in a total parameter count of 0.76 billion for OMPGPT, making it more compact than our baseline models.



\noindent
\textbf{OMPGPT Inference}. We convert the OpenMP pragma generation problem into a code generation task by replacing the OpenMP pragmas in training.
Figure~\ref{fig:prompt_base} (a) illustrates inference with OMPGPT for code generation. Originating from a GPT-based model, OMPGPT inherits the capability for generation tasks given input code. OMPGPT is trained on pre-processed OpenMP code where pragmas are moved to the end of loops. When prompted with a prefix like \texttt{\#pragma omp}, as shown in Figure~\ref{fig:prompt_base} (b), OMPGPT demonstrates its specialized ability to generate relevant OpenMP pragmas. This highlights OMPGPT's aptitude for intuitively continuing code sequences tailored to the syntax and structure of OpenMP directives, showcasing practical utility for real-world programming scenarios.




\begin{figure}[ht]
\centering
\includegraphics[width=\textwidth]{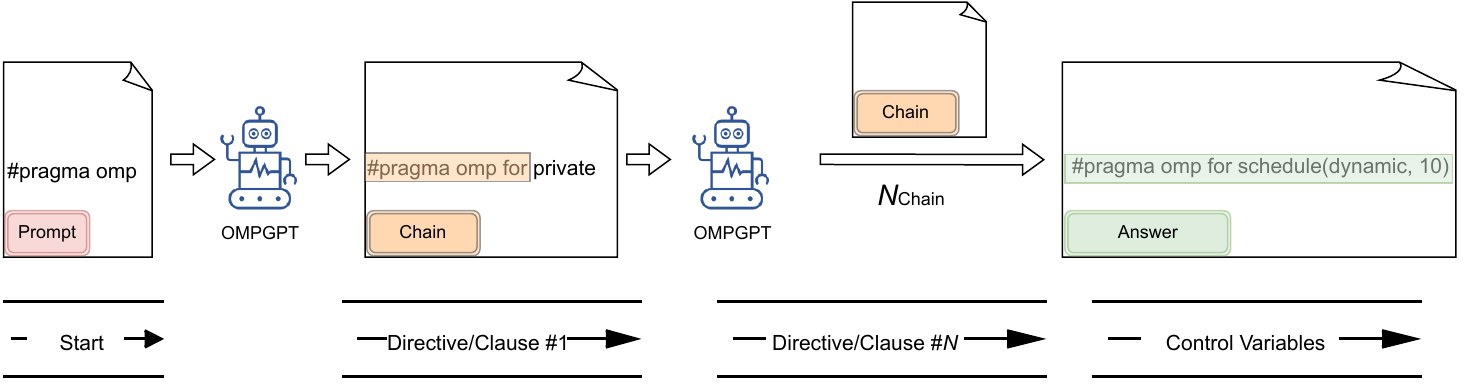} 
\caption{Chain-of-OMP. Start phase: using the standard OpenMP pragma generation prompts for OMPGPT. Directive/clause generation phase: the first generated Directive/clause is augmented to the previous input as the input for the next chain. This phase is updated till it number of chains reaches parameter $N_{\text{chain}}$. The final Control variable generation phase generates the last component to complete the chain-of-OMP and generate the OpenMP pragma.}
\label{fig:chain}
\end{figure}

\subsection{Chain-of-OMP}
Chain-of-thought (CoT)~\cite{wei2023chainofthought} facilitates complex reasoning in NLP tasks by guiding LLMs through intermediate steps. Inspired by the "think step by step" methodology, we developed Chain-of-OMP, a novel prompt engineering technique to enhance OMPGPT in generating OpenMP pragmas.

We consider an OpenMP pragma as comprising three major components:
\begin{itemize}
    \item Directive Prefix (\texttt{\#pragma omp}): 
    This universal prefix signals the presence of an OpenMP directive to the compiler. Within our model, this prefix is used in prompts to orient OMPGPT towards generating OpenMP pragmas, steering it away from generic code generation. We represent it as \texttt{<prefix>}.
    \item Directives and Clauses. This segment outlines the type of parallel or work-sharing construct being utilized and details its specific behavior. We represent them as \texttt{<dc>}.
    \item Control Structure. A control structure typically immediately follows the directives. It is essential for the behavior of directives like \texttt{\#omp parallel reduction}. We represent it as \texttt{<cs>}.
\end{itemize}


In OpenMP syntax, these three components are listed sequentially (i.e., \texttt{<prefix><dc><cs>}), perfectly fitting the instinct of LLMs. Leveraging this, Chain-of-OMP operates as an automated sequence of prompts, as depicted in Figure~\ref{fig:chain}.  The process begins with standard OpenMP pragma generation prompts (i.e., \texttt{<prefix>}). The output from the first stage is then refined, retaining only the generated directive/clause, which is subsequently passed to the next OMPGPT client to generate the following component. In other words, we only retain \texttt{<prefix><dc>} part of the generated output and ignore the rest. This retained part is then fed as the prompt for the next stage. In general, the components of an OpenMP pragma expand incrementally, with one element added at each inference stage, until it reaches a user-defined limit, $N_{\text{chain}}$. The input $I_n$ for a chain component chain$_n$ is defined by Equation~\ref{eq:chain}.
\begin{equation}
\label{eq:chain}
    \text{I}_n = concat(\text{I}_{n-1}, \text{first generated directive/clause in O}_{n-1}  )
\end{equation}
where $I_n$, $O_n$   stands for the input and output of chain$_n$. $concat$ stands for the concatenate operation. OpenMP pragmas have a different number of components. Consequently, we need a different number of chains to generate the complete OpenMP pragma. We let the user set the maximum number of chains limit by setting $N_{\text{chain}}$. The default value of $N_{\text{chain}}$ is 256 to avoid an infinity loop. The value aligns with the commonly used maximum output length of LLMs.
The design of Chain-of-OMP has the following advantages:
\begin{itemize}
    \item Mimicking Expert User Inquiry: This approach closely replicates the querying process employed by experienced OpenMP users. Instead of requesting a complete pragma with a basic prompt (using just \texttt{\#pragma omp}), skilled users often provide more specific information, such as directives (e.g., \texttt{\#pragma omp for}), to refine their inquiry about the remaining parts of the pragma. Chain-of-OMP embodies this nuanced approach, leading to more targeted and accurate pragma generation.
    \item Enhancing Performance Across Various LLMs: By selectively retaining the initial directive/clauses and discarding the rest, Chain-of-OMP effectively narrows the search space for subsequent chains. This approach is akin to the strategy in the classic Monty Hall problem, where a "pick again" method theoretically increases the chances of accuracy. Such a strategy is anticipated to improve OpenMP pragma generation accuracy not just for OMPGPT but for other LLMs as well.
    \item Automation of the Process: A key strength of Chain-of-OMP is its fully automated chain generation capability. This contrasts traditional chain-of-thought techniques, where users typically need to craft prompts for each step manually. Users usually do not need to specify the value of $N_{\text{chain}}$, as the model will stop when it predicts there is an end of a pragma. Explicitly defining $N_{\text{chain}}$ is also an option for expert users who want the model to run $n$ chain stages. The automation in Chain-of-OMP streamlines the process, making it more efficient and user-friendly.

\end{itemize}



\subsection{Fine-tuning}
\label{sec: finetune}
Due to the complexity of OpenMP pragmas, previous works either train an ML model or fine-tune a language model for specific tasks covering a limited selection of pragmas.  In our work, we have employed a strategic fine-tuning approach to demonstrate the enhanced performance of OMPGPT after fine-tuning and to facilitate a comprehensive comparison with baseline models.

We fine-tune the pre-trained OMPGPT model using the AdamW optimizer~\cite{loshchilov2017decoupled}, a variant of the Adam optimizer known for its effectiveness in large models and datasets. This process incorporates a linear warm-up phase over the initial 100 steps. The warm-up phase gradually increases the learning rate from zero to the initial learning rate set for training, helping to stabilize the model's learning process in its early stages. Following the warm-up, we implement a linear decay in the learning rate for the remaining steps. This decay approach gradually reduces the learning rate, allowing for finer adjustments to the model's weights as it converges toward optimal performance.

\section{Evaluation} 
\label{sec-evaluation}




\subsection{Model Perplexity}
\noindent
\textbf{Task Definition}. This subsection evaluates the general knowledge of OpenMP code possessed by the base model. We assess its ability to generate code using perplexity score, a common metric in language processing. Perplexity essentially measures how surprised a model is by the next word in a sequence. Lower perplexity indicates the model can better predict upcoming elements and thus has a better understanding of the language. In this context, we use perplexity to gauge the model's grasp of OpenMP code structure.

\noindent
\textbf{Evaluation Setup}. As described in Section~\ref{sec:training}, we use 10\% of the HPCorpus dataset as a test set unknown to OMPGPT. We calculate OMPGPT's perplexity (PP) using the following Equation~\ref{eq: per}.
\begin{equation}
\label{eq: per}
    PP = \exp\left(-\frac{1}{N}\sum_{i=1}^{N} \log P(w_i|w_1, w_2, \ldots, w_{i-1})\right)
\end{equation}
Where $P(w_i, w_1, w_2, \ldots, w_{i-1})$ is the probability of the first word $w_i$ occurring, given the sequence of the subsequent words $w_1, w_2, \ldots, w_{i-1}$. We use prompt (a) in Figure~\ref{fig:prompt_base} for general code generation and calculate the perplexity. We compare the perplexity result of OMPGPT with MonoCoder and other open-source language models.

\noindent
\textbf{Baselines}. Table~\ref{tab: perplexity} compares perplexity scores on OpenMP code for various
\begin{wraptable}{r}{0.45\textwidth}
\centering
\caption{Perplexity Comparison across Open-source Language Models.}
\label{tab: perplexity}
\resizebox{0.35\textwidth}{!}{%
\begin{tabular}{cccc}
\hline
Model     & Size (B) & C & C++ \\ \hline
OMPGPT    & \textbf{0.76}     & 3.54           & 3.66             \\
MonoCoder & 0.89     & 3.51           & 3.69             \\
PolyCoder & 2.7      & 2.33           & 2.99             \\
GPT-Neo   & 2.7      & 3.69           & 2.87             \\
GPT-J     & 6        & 2.82           & 2.47             \\
CodeX     & 12       & 2.55           & 1.95             \\
StarCoder & 15.5     & 1.71           & 2.01             \\
GPT-NeoX  & 20       & 2.37           & 2.32             \\ \hline
\end{tabular}}
\end{wraptable}
 open-source code LLMs. These models have different parameter sizes and were all trained with extensive parallel learning (PL) data, enabling their code generation capabilities. However, most of them are generic models and are included here for perplexity evaluation only. MonoCoder stands out as the only model specifically focused on OpenMP. Additionally, it was trained and tested with the HPCorpus dataset, making it a well-suited baseline for our later investigation of OpenMP pragma generation.

\noindent
\textbf{Results}. As shown in Table~\ref{tab: perplexity}, OMPGPT achieves competitive scores (3.54 for C and 3.66 for C++) despite its smaller size (0.76B) compared to models like PolyCoder (2.7B) or GPT-J (6B). While larger models like StarCoder (15.5B) generally achieve lower perplexity, OMPGPT demonstrates efficiency by competently bridging this gap with larger models. 
 This trend suggests a correlation between model size and perplexity, but the efficiency of OMPGPT is notable, given its significantly smaller size. It competently bridges the gap with larger models, indicating a promising direction for efficient model design in the code generation tasks.


\subsection{OpenMP Pragma Generation with OMPGPT Base Model}
\label{sec:eval_base}

\noindent
\textbf{Task Definition}. OMPGPT is trained with the preprocessed OpenMP code in HPCorpus.  The capability of generating any OpenMP pragma distinguishes OMPGPT from most existing code-oriented Large Language Models (LLMs). This task evaluates OMPGPT's performance on the 15 most prevalent OpenMP pragmas found in the HPCorpus test set to assess this capability.

\noindent
\textbf{Evaluation Setup}. We followed the Prompt (b) in Figure~\ref{fig:prompt_base}, where the input code is followed by the hint \texttt{\#pragma omp}  to signal to OMPGPT that we are performing the pragma prediction instead of generic code generation. 
Evaluation metrics focus on accuracy, specifically gauging the congruence between OMPGPT's outputs and the original OpenMP pragmas.

We employed a strict matching criterion, wherein a generated pragma is considered correct only if it precisely matches the corresponding pragma in the test dataset. For instance, although the following two OpenMP pragmas are functionally equivalent, they are not deemed correct under this strict criterion:

{
\texttt{\#pragma omp parallel for reduction(+:sum) private(var)}
}

{
\texttt{\#pragma omp parallel for private(var) reduction(+:sum)}
}

\begin{figure}[t]
\centering
\includegraphics[width=\textwidth]{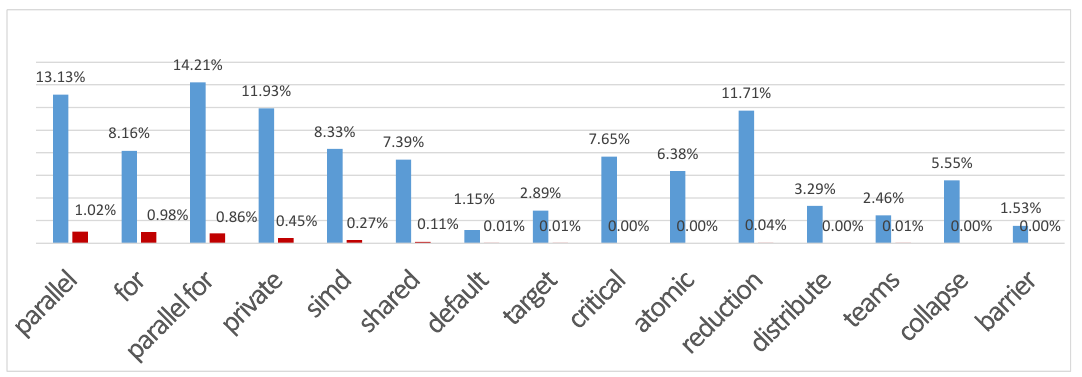} 
\caption{Pragma Generation Accuracy. Blue: OMPGPT. Red: GPT-Neo.}
\label{fig:base_res}
\end{figure}

\noindent
\textbf{Results}. Figure~\ref{fig:base_res} compares OMPGPT with its base model GPT-Neo. Notably, OMPGPT outperforms GPT-Neo, indicating the performance gain is achieved as OMPGPT successfully learned the task of OpenMP pragma generation. However, as shown in the figure, the strict matching criterion results in lower scores, highlighting the challenge of achieving perfect accuracy in this task.

\vspace{-0.1cm}

\subsection{OpenMP Pragma Generation using Chain-of-OMP}
\label{sec:chain_eval}

\textbf{Task Definition}. 
As shown in Section~\ref{sec:eval_base}, OMPGPT's base model can generate OpenMP pragmas, but accuracy is limited due to the complexities of OpenMP syntax. This section evaluates our proposed Chain-of-OMP method, assessing its effectiveness in improving pragma generation compared to basic prompts. This comparison will highlight Chain-of-OMP's ability to address these syntactic challenges.

\noindent
\textbf{Evaluation Setup}. We evaluate chain-of-OMP for four specific OpenMP clauses: \texttt{for schedule}, \texttt{collapse}, \texttt{teams}, and \texttt{target}. These pragmas are well-aligned with the three-component structure outlined earlier. We compare Chain-of-OMP's accuracy with the base prompt to quantify its performance improvement.

\noindent
\textbf{Results}. Table~\ref{tab:chain_res} showcases the accuracy gains achieved by Chain-of-OMP.
\begin{wraptable}{r}{0.52\textwidth}
\centering
\caption{Chain-of-OMP effectiveness evaluation with OMPGPT.}
\label{tab:chain_res}
\resizebox{0.45\textwidth}{!}{%
\begin{tabular}{cccc}
\hline
              & Basic Prompt & Chain-of-OMP  \\ \hline
for schedule & 1.3\%      & 7.8\%          \\
collapse & 0.1\%      & 2.4\%          \\
teams & 0.1\%      & 0.3\%          \\
target        & 2.9\%      & 16.9\%        \\ \hline
\end{tabular}}
\end{wraptable}
For \texttt{for schedule}, Chain-of-OMP significantly improves accuracy from 1.3\% to 7.8\%, representing a substantial 600\% increase. Similar improvements are observed for \texttt{collapse}  (0.1\% to 2.4\%, a 2300\% increase) and \texttt{target} (2.9\% to 16.9\%, a 580\% increase). The smaller improvement for \texttt{teams} (0.1\% to 0.3\%) is due to limited training data for this specific pragma.

\subsection{Fine-tuning OMPGPT for Specific Pragma Generation}
\textbf{Task Definition}. As discussed in Section \ref{sec: finetune}, fine-tuning is necessary due to the complexity of OpenMP pragmas. We fine-tuned OMPGPT for three sub-tasks: private, reduction, and SIMD. We evaluate its performance against two baselines: MonoCoder (fine-tuned for pragma generation) and GPT-3.5 (not fine-tuned due to time/cost constraints). We also applied Chain-of-OMP with the fine-tuned OMPGPT for each sub-task, setting the number of chains $N_{\text{chain}}$ to 2 based on the evaluation clauses' structure.

\noindent
\textbf{Evaluation Setup}. We used the test set from our fine-tuning dataset for each sub-task. We employed precision, recall, F1-score, and accuracy metrics. The evaluation consisted of two subtests:
\begin{enumerate}
    \item Subtest 1 (Clause Matching): This subtest checks if the predicted clause matches the expected clause.
    \item Subtest 2 (Stricter Matching): This stricter subtest requires both the predicted clause and its control structure to match the expected ones.
\end{enumerate}

For both GPT-3.5 and Monocoder, we used the same test set employed for OMPGPT to ensure a fair comparison. Results for MonoCoder are reported from its original paper since it uses the same dataset splits as OMPGPT. We note that some entries in Tables~\ref{tab:fine_tune_test1} and~\ref{tab:fine_tune_test2} were unavailable for MonoCoder since they were not reported in their paper.



\begin{table}[!h]
\centering
\caption{Fine-tuning results for \texttt{private}, \texttt{reduction}, and \texttt{SIMD} on the first test (Subset 1). P = Precision, R = Recall, F1 = F1 Score, Acc = Accuracy. Note Monocoder does not support SIMD.}
\label{tab:fine_tune_test1}
\begin{tabular}{l|rrrr|rrrr|rrrr}
\hline
\textbf{Model} & \multicolumn{4}{|c|}{\texttt{private}} & \multicolumn{4}{|c|}{\texttt{reduction}} & \multicolumn{4}{|c}{\texttt{SIMD}} \\
\hline
          & P & R & F1 & Acc & P & R & F1 & Acc & P & R & F1 & Acc\\ \hline
OMPGPT    & 0.94 & 0.91 & 0.92 & 0.96 & 0.92 & 0.91 & 0.915 & 0.98 & 0.88 & 0.82   & 0.85 & 0.93 \\
MonoCoder & 0.89 & 0.83 & 0.86 & 0.94 & 0.99      & 0.81   & 0.891 & 0.98 &  -       & -    & -  & -   \\
GPT-3.5    & 0.4  & 0.16 & 0.23 & 0.77 & 0.53      & 0.57   & 0.549 & 0.92 & 0.51  & 0.55   & 0.53 & 0.84  \\ \hline
\end{tabular}
\end{table}


\noindent
\textbf{Results}. Tables~\ref{tab:fine_tune_test1} and~\ref{tab:fine_tune_test2} show the performance on the two subtests for private, reduction, and SIMD clauses (MonoCoder doesn't support SIMD). Overall, OMPGPT consistently outperforms both MonoCoder and GPT-3.5 models in the two subtests. As expected, the performance drops for all models in the second subtest (clause and control structure). However, notably, the Chain-of-OMP prompt improves the accuracy of OMPGPT in predicting both the clause and control structure correctly in the second subtest.


\begin{table}[!h]
\centering
\caption{Fine-tuning results for \texttt{private}, \texttt{reduction}, and \texttt{SIMD} on the second test (Susbet 2). Note Monocoder does not support SIMD.}
\label{tab:fine_tune_test2}

\begin{tabular}{l|rrrr|rrrr|rrrr}
\hline
\textbf{Model} & \multicolumn{4}{|c|}{\texttt{private}} & \multicolumn{4}{|c|}{\texttt{reduction}} & \multicolumn{4}{|c}{\texttt{SIMD}} \\
\hline
          & P & R & F1 & Acc & P & R & F1 & Acc & P & R & F1 & Acc\\ \hline
OMPGPT    & 0.74      & 0.61   & 0.67 & 0.51     & 0.78       & 0.71    & 0.74 & 0.57     & 0.64      & 0.69   & 0.66 & 0.42 \\
Chain-of-OMP     & 0.76      & 0.77   & 0.76 & 0.64     & 0.77       & 0.78    & 0.77 & 0.60     & 0.66      & 0.73   & 0.69 & 0.58 \\
MonoCoder & -       & -    & -  & 0.48     & -        & -     & -  & 0.52     & -       & -    & -  & -   \\
GPT-3.5    & 0.23      & 0.34   & 0.27 & 0.14     & 0.71       & 0.55    & 0.62 & 0.51     & 0.61      & 0.55   & 0.58 & 0.40  \\ \hline
\end{tabular}
\end{table}

The effectiveness of Chain-of-OMP is showcased by analyzing two selected OpenMP pragmas, as detailed in Section~\ref{sec:chain_eval}. Furthermore, we extended our evaluation to include a smaller set comprising ten samples of the top 15 most common pragmas in this test. This testing led to two significant observations:

\begin{itemize}
    \item Consistent Performance Enhancement: Chain-of-OMP consistently maintained or improved performance across all test sets. In no instance did its application lead to a performance decline.
    \item Notable Improvement in Majority of Tasks: Impressively, Chain-of-OMP improved performance in 11 out of the 15 pragma generation tasks, demonstrating its substantial efficacy in enhancing OpenMP pragma generation across various tasks.
\end{itemize}

\vspace{-10pt}

\section{Related Work} 
Recent works have explored language models (LMs) for HPC tasks as described in~\cite{kadosh2023domain}.  Studies have fine-tuned existing code LLMs for tasks such as race detection~\cite{dingHPCGPTIntegratingLarge2023} and generating parallel code like MPI routines~\cite{nichols2023modeling}. The latter work also explored the task of generating accurate OpenMP pragmas. In contrast, \cite{kadoshAdvisingOpenMPParallelization2023} formulated predicting OpenMP pragmas as a multi-label classification task. 

One of the most recent studies utilizing LLMs for OpenMP pragma generation is MonoCoder~\cite{kadosh2023domain}. This work fine-tunes a domain-specific language model on the HPCorpus dataset specifically for OpenMP pragma generation. OMPGPT stands out from MonoCoder due to its novel training approach. By strategically repositioning OpenMP pragmas during training, OMPGPT aligns with the left-to-right processing inherent to GPT models. This approach, along with the proposed Chain-of-OMP prompting technique, leads to consistent performance improvements over MonoCoder.

Chain-of-thought prompting improves LM reasoning ability, showing promise on math word problems, commonsense reasoning~\cite{wei2023chainofthought}, and summarizing software components~\cite{rukmonoAchievingHighLevelSoftware2023}. This prompting strategy has also enhanced code generation from models like ChatGPT~\cite{liStructuredChainofThoughtPrompting2023,liuImprovingChatGPTPrompt2023} and bug reproduction from reports~\cite{fengPromptingAllYou2023}. Most relevantly, it has been applied to correcting compilable code~\cite{chenCompCodeVetCompilerguidedValidation2023}.




\label{sec-related-work}

\section{Conclusion and Future Work} 
While large language models have transformed natural language processing.  they struggle with domain-specific problems like HPC tasks. This work presents OMPGPT, a 0.76B domain-specific model trained for OpenMP pragma generation on HPC data. Our evaluation shows that OMPGPT outperforms larger LLMs like GPT-3.5. Moreover, our novel chain-of-OMP prompting technique assists OMPGPT in improving accuracy through step-by-step prompts. Our findings suggest that smaller LLMs can achieve excellent performance on specific tasks with proper domain training and prompting techniques. This paves the way for more accessible and efficient LLMs.

Future work includes evaluating Chain-of-OMP on other OpenMP clauses, extending context windows beyond loop snippets, training LLMs from scratch on our data with Chain-of-OMP prompting, and exploring this approach for other HPC tasks \cite{arijit}.

\label{sec-conclusion}

\section{Acknowledgement}
This project was funded by NSF (\#$2211982$) and Intel Labs. We would like to thank them for their generous support. 
Additionally, we extend our gratitude to the Research IT team\footnote{https://researchit.las.iastate.edu/} of Iowa State University for their continuous support in providing access to HPC clusters for conducting the experiments of this project.

\bibliographystyle{splncs04}

\bibliography{main.bib}

\end{document}